\documentclass[twocolumn]{aastex63}
\submitjournal{BAAS}
\usepackage{CJKutf8, xcolor}

\usepackage[hyphens]{url}
\usepackage{hyperref}
\usepackage[hyphenbreaks]{breakurl}
\usepackage{tikz}
\usetikzlibrary{positioning, shapes}

\shorttitle{Best Practices for ML in Astronomy}
\shortauthors{Huppenkothen, Ntampaka et al.}

\begin{document}


\author[0000-0002-1169-7486]{Daniela Huppenkothen}
\altaffiliation{Both authors contributed equally to this work.}
\affiliation{SRON Netherlands Institute for Space Research, Niels Bohrweg 4, 2333CA Leiden, The Netherlands}
\affiliation{Anton Pannekoek Institute for Astronomy, University of Amsterdam, Science Park 904, 1098 XH, Amsterdam, The Netherlands}

\author[0000-0002-0144-387X]{Michelle Ntampaka}
\altaffiliation{Both authors contributed equally to this work.}
\affiliation{Space Telescope Science Institute, Baltimore, MD 21218,USA}
\affiliation{Department of Physics \& Astronomy, Johns Hopkins University, Baltimore, MD 21218,USA}
\email{d.huppenkothen@sron.nl, mntampaka@stsci.edu}

\author[0000-0003-3207-8868]{Matthew Ho}
\affiliation{CNRS \& Sorbonne Universit\'{e}, Institut d’Astrophysique de Paris (IAP),
UMR 7095, 98 bis bd Arago, F-75014 Paris, France}

\author[0000-0001-9256-5516]{Morgan Fouesneau}
\affiliation{Max Plank Institute for Astronomy (MPIA), Königstuhl 17, D-69117 Heidelberg, Germany}

\author[0000-0001-6706-8972]{Brian Nord}
\affiliation{Fermi National Accelerator Laboratory, P. O. Box 500, Batavia, IL 60510, USA}
\affiliation{Kavli Institute for Cosmological Physics, University of Chicago, Chicago, IL 60637, USA}
\affiliation{Department of Astronomy and Astrophysics, University of Chicago, Chicago, IL 60637, USA}

\author[0000-0003-4797-7030]{J.E.G.~Peek}
\affiliation{Space Telescope Science Institute, Baltimore, MD 21218,USA}
\affiliation{Department of Physics \& Astronomy, Johns Hopkins University, Baltimore, MD 21218,USA}

\author[0000-0002-6408-4181]{Mike Walmsley}
\affiliation{Jodrell Bank Centre for Astrophysics, Department of Physics \& Astronomy, University of Manchester, Manchester, M13 9PL, UK}
\affiliation{Dunlap Institute for Astronomy \& Astrophysics, University of Toronto, 50 St. George Street, Toronto, ON M5S 3H4, Canada}

\author[0000-0002-5077-881X]{John F. Wu}
\affiliation{Space Telescope Science Institute, Baltimore, MD 21218,USA}
\affiliation{Department of Physics \& Astronomy, Johns Hopkins University, Baltimore, MD 21218,USA}


\author[0000-0001-8868-0810]{C.~Avestruz}
\affiliation{Leinweber Center for Theoretical Physics, University of Michigan, Ann Arbor, MI 48109, USA}
\affiliation{Department of Physics, University of Michigan, Ann Arbor, MI 48109, USA}

\author[0000-0003-2027-399X]{Tobias Buck}
\affiliation{Universit\"at Heidelberg, Interdisziplin\"ares Zentrum f\"ur Wissenschaftliches Rechnen, Im Neuenheimer Feld 205, 69120 Heidelberg, Germany}
\affiliation{Universit\"at Heidelberg, Zentrum f\"ur Astronomie, Institut f\"ur Theoretische Astrophysik, Albert-Ueberle-Straße 2, 69120 Heidelberg, Germany}

\author{Massimo Brescia}
\affiliation{Department of Physics ``E. Pancini,'' University Federico II of Napoli, Via Cinthia 21, I-80126 Napoli, Italy}
\affiliation{INAF Astronomical Observatory of Capodimonte, Salita Moiariello 16, I-80131 Napoli, Italy}

\author[0000-0003-2808-275X]{Douglas P. Finkbeiner}
\affiliation{Department of Physics, Harvard University, 17 Oxford St., Cambridge, MA 02138, USA}
\affiliation{Harvard-Smithsonian Center for Astrophysics, 60 Garden St., Cambridge, MA 02138, USA}

\author[0000-0003-4700-663X]{Andy D. Goulding}
\affiliation{Department of Astrophysical Sciences, Princeton University, Princeton, NJ 08544, USA}

\author[0000-0001-5570-7503]{T.~Kacprzak}
\affiliation{Swiss Data Science Center, Paul Scherrer Institute, 5303 Villigen, Switzerland}

\author[0000-0002-8873-5065]{Peter Melchior}
\affil{Department of Astrophysical Sciences, Princeton University, Princeton, NJ 08544, USA}
\affil{Center for Statistics \& Machine Learning, Princeton University, Princeton, NJ 08544, USA}

\author[0000-0003-3784-5245]{Mario Pasquato} \affil{Département de Physique, Universite de Montréal, Montreal, Quebec H3T 1J4, Canada} \affil{Physics and Astronomy Department Galileo Galilei, University of Padova, Vicolo dell’Osservatorio 3, I–35122, Padova\\} \affil{Mila - Quebec Artificial Intelligence Institute, Montreal, Quebec, Canada} \affil{Ciela, Computation and Astrophysical Data Analysis Institute, Montreal, Quebec, Canada}

\author[0000-0001-7772-0346]{Nesar Ramachandra}
\affiliation{Computational Science Division, Argonne National Laboratory, Lemont, IL, USA}
\affiliation{High Energy Physics Division, Argonne National Laboratory, Lemont, IL, USA}

\author[0000-0001-5082-9536]{Yuan-Sen Ting}
\affiliation{School of Computing, Australian National University, Acton, ACT 2601, Australia}
\affiliation{Research School of Astronomy \& Astrophysics, Australian National University, Cotter Rd., Weston, ACT 2611, Australia}
\affiliation{Department of Astronomy, The Ohio State University, Columbus, USA}

\author{Glenn van de Ven}
\affil{Department of Astrophysics, University of Vienna, T\"urkenschanzstra{\ss}e 17, 1180 Vienna, Austria}

\author{Soledad Villar}
\affil{Department of Applied Mathematics and Statistics, Johns Hopkins University, Baltimore, MD, USA}
\affil{Mathematical Institute for Data Science, Johns Hopkins University, Baltimore, MD, USA}

\author[0000-0002-5814-4061]{V.A.~Villar}
\affil{Center for Astrophysics \textbar{} Harvard \& Smithsonian, 60 Garden Street, Cambridge, MA 02138-1516, USA}

\author[0000-0002-6316-3996]{Elad Zinger}
\affiliation{Centre for Astrophysics and Planetary Science, Racah Institute of Physics, The Hebrew University, Jerusalem 91904, Israel}

\title{Constructing Impactful Machine Learning Research for Astronomy:\\ Best Practices for Researchers and Reviewers} 

\begin{abstract}

Machine learning has rapidly become a tool of choice for the astronomical community. It is being applied across a wide range of wavelengths and problems, from the classification of transients to neural network emulators of cosmological simulations, and is shifting paradigms about how we generate and report scientific results. At the same time, this class of method comes with its own set of best practices, challenges, and drawbacks, which, at present, are often reported on incompletely in the astrophysical literature. With this paper, we aim to provide a primer to the astronomical community, including authors, reviewers, and editors, on how to implement machine learning models and report their results in a way that ensures the accuracy of the results, reproducibility of the findings, and usefulness of the method.\\
\end{abstract}

\section{Introduction}

\begin{figure*}
    \centering
    \includegraphics[width=0.7\textwidth]{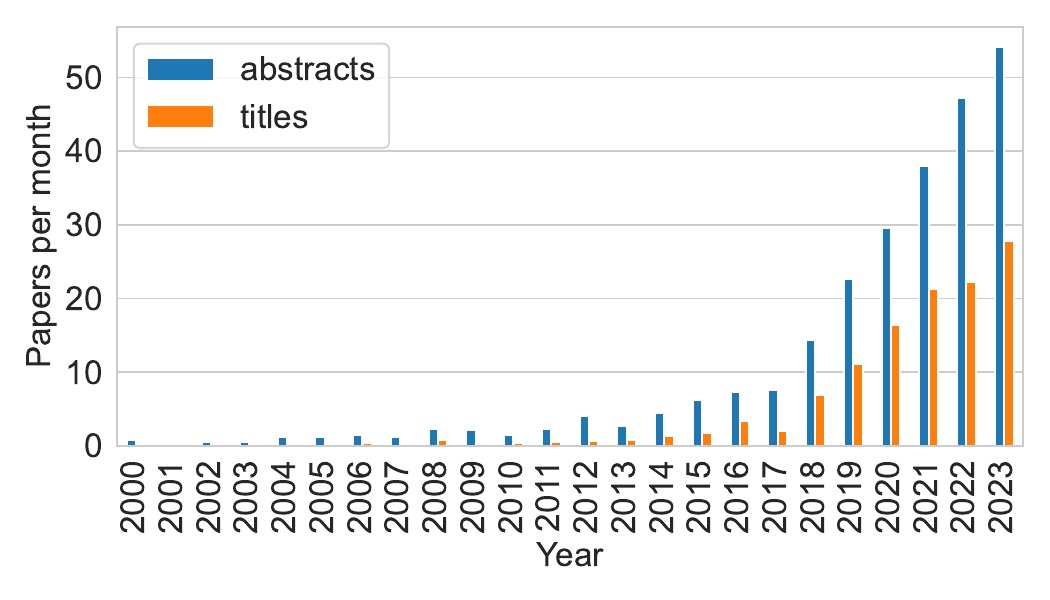}
    \caption{Number of refereed publications per month that include the terms ``machine learning'' or ``artificial intelligence'' in the title (orange) or abstract (blue) in the top 25 astronomy and astrophysics-related journals.}
    \label{fig:pubs}
\end{figure*}

Data-driven astronomy is experiencing a period of unprecedented growth. Large volumes of observational data, data sets with complex observational biases, and realistic high-resolution simulations have created a perfect environment for modern data analysis tools, such as machine learning, to flourish.  Peer-reviewed machine learning astronomy research is growing at an astonishing rate, with a publication doubling time of about eighteen months (Figure \ref{fig:pubs}).  

As machine learning\footnote{Some communities favor ``Artificial Intelligence'' (AI) as a term for this class of methods. In this manuscript, we limit ourselves to the term ``machine learning'' for clarity, but it should be understood that we mean to encompass both.} (ML) methods become standard tools in the astronomical community, there is an increasing need for best practices for how to build and implement these methods, how to apply them to astronomical data sets, and how to report the results to the community. This is partly driven by the idiosyncrasies of astronomical data; challenges in the field include heterogeneous errors and a dearth of ground-truth training data. These challenges and others can often make astronomical data analysis problems mismatched to standard approaches or at least require careful validation of the methods used. 

This paper was motivated by the authors' experiences in developing methods and doing research at the interface of astronomy and machine learning across a wide range of astrophysical problems. Our goal with this manuscript is not to be prescriptive, but to provide suggestions to ensure the use of machine learning on astronomical data meets three criteria: it should be accurate, reproducible, and useful. Machine learning as applied to astronomical data should provide \textit{accurate} results in the sense that reported results are ideally not biased, for example due to a mismatch between training and target data. Results should be \textit{reproducible} in that they should be stable across repeat training/prediction cycles, and that the manuscript should include all relevant details for other groups to reproduce the results. Finally, results should be \textit{useful} in that they ought to provide some advantage compared to established alternative approaches. 

In Section \ref{sec:rules}, we provide six suggestions for machine learning research and associated manuscripts in astronomy. 
These are meant to provide a starting point and highlight particularly relevant best practices that we urge researchers to consider throughout their research. They are not meant to be seen as absolute: rules can (and should) be broken when there is a good reason to do so (Section \ref{sec:breaking}). Thus, we suggest their application with thoughtfulness and nuance when carrying out ML research projects and when reviewing them. 

We are also acutely aware that ML is a particularly fast-paced field, and so this document should be considered a starting point for a wider, evolving conversation around best practices. The astronomical community, as a whole, is rapidly adopting increasingly sophisticated ML models and improving how these models are being embedded in research projects. Most of the work in this space is both ground-breaking and interdisciplinary, and we believe the rapid pace of adoption merits the higher-level view upon the use of these methods presented here. In this context, we readily admit that many of the works by the authors of this paper are certain to fall short in one or more aspects of the rules presented below. The discussions that have led to this manuscript have, in themselves, be immensely valuable in clarifying the components of a thoughtful, nuanced approach to ML research and applications in astronomy.

\section{The Rules}
\label{sec:rules}

In this section, we describe six important considerations for designing and implementing machine learning tools for astronomical data, and for reporting the results. Many astronomical problems are not well-suited for black-box applications of algorithms and software, and careful consideration is warranted in evaluating and interpreting results. Some of the rules we describe below are similar to general best practices for research and are already widely adopted throughout the community. Nevertheless, we think it is important to spell out these practices as a starting point for discussion among ML experts and as guidance to researchers starting out applying ML to their own astronomical challenges. 

\begin{figure*}
    \centering
    \includegraphics[width=\textwidth]{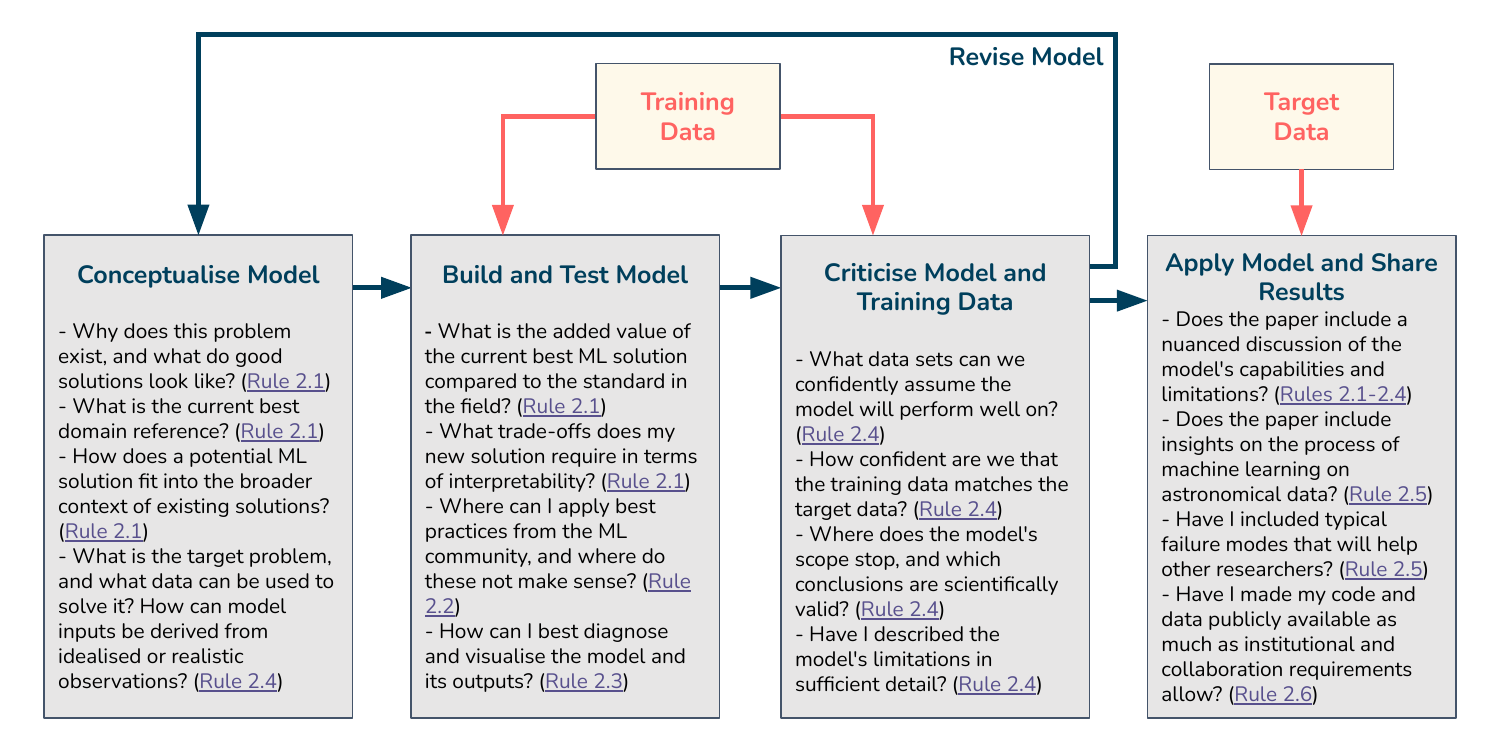}
    \caption{Box's loop for ML in astronomy.}
    \label{fig:boxloop}
\end{figure*}

In Figure \ref{fig:boxloop}, we present an overview of the rules and key questions that researchers might ask themselves throughout a project's life cycle. The schematic borrows from George Box's notion of research as an iterative process \citep{box1976science} and the interpretation by \citet{blei2014} in the context of exploratory data analysis and latent variable models. In this view, critiquing and revising statistical models is a key component of scientific practice, and we suggest a similar approach to building and validating ML models. This Figure is intended for researchers working on a project, and in Appendix \ref{sec:appendix} we present an accompanying quick-start guide for researchers refereeing relevant manuscripts or proposals.

\subsection{Compare against a domain reference and put results into the larger context} \label{sec:2.1}
 
The goal of most applications of machine learning in astronomy is to infer astrophysical knowledge. Major motivations for implementing machine learning methods include (1) decreasing the computational time and costs associated with scalability to larger data sets and samples; (2) improving the robustness or precision of solutions to astrophysical questions; (3) facilitating model sharing and automation to make collaboration easier; (4) implementing machine learning models for problems where no other solution exists. Because scientific progress is inherently an incremental process where new results are built upon previous findings and methods, all but the last motivation imply that a corpus of work exists on the topic at hand. 

Explicitly evaluating new ML methods against traditional ones is vital to establish trust within the astronomical community. This might start even before beginning a particular project, as part of project scoping, by evaluating the following questions: Why does this problem exist, and what do good solutions look like? How would an ML solution fit into the broader context of existing solutions? How do its expected strengths and weaknesses compare to alternatives? What would someone need to do to put this into practice? While an evaluation of ethical and privacy concerns is rarely required in the context of projects involving astronomical data, we note that this is not the case in projects and fields where ML outputs are used for decision-making that affects human lives. In these situations, an impact assessment focused on the ethical and privacy implications should be at the heart of project development.  

During the course of the research, results derived with a machine learning model should be compared to the current standards in the field by considering what value is added by deploying a machine learning model. This is an especially important consideration in the context of interpretability. An ML model may be less interpretable than the methods it is being compared to, especially when those methods are motivated by the underlying physics of the problem. It is therefore important to discuss how a new ML model improves on the current best approach while also considering scenarios where a more traditional method might be a more appropriate solution, e.g.~because interpretability is particular important for a specific application or because the new model underperforms for specific data sets or use cases. Researchers should clearly and quantitatively highlight the benefits of the new model within the scope of the larger astrophysical question and the context of alternative approaches to the problem.  

The relevant context may include not only the specific problem at hand but also how the ML approach might facilitate (or introduce additional bias into) larger-scale analyses that use the outputs of the ML model, e.g.~in the context of population inference or cosmology. For example, ML might be used to estimate the masses of galaxy clusters in the context of estimating the Halo Mass Function (HMF). One might then ask how a particular improvement in estimated cluster masses affects the survey estimation of the HMF? And how would the researchers ensure that the cluster and galaxy selection function in the training set is equal to that of the survey?

As a concrete example for an application of the concepts presented in this rule, \citet{matzeu2022.515.6172M} present a neural network surrogate model for a 2.5D radiative transfer model calculating the X-ray emission from accretion disk winds. In a traditional approach, computing an X-ray spectrum from a set of astrophysical parameters and initial conditions requires several hours of computing time. The surrogate model presented in the paper can do the same in a fraction of a second. They compare their emulator to the current standard model (a fast model based on grid-based linear interpolation) on both simulated data and \textit{XMM-Newton} observations. This comparison pits the surrogate model against the current standard approach within the context of an end-to-end analysis, and includes the comparison of full posterior distributions using both models. This paper's in-depth discussion is a good example of the concepts advocated in this rule: it includes a clear problem statement and motivation for implementing a machine learning solution, a comparison with the standard approach on both simulated and real data, and a discussion of the advantages and shortcomings of the new approach.

To summarize, machine learning should not be used for machine learning's sake; the reason for implementing an ML solution should be clearly and prominently articulated. To ensure that the research findings are useful to the broader astronomical community, the advantages of the technique should be highlighted, and the results should be carefully compared against the work flow and performance of a conventional method.

\subsection{Adopt best practices from the ML community} \label{sec:2.2}

Though ``machine learning'' as a term is often reduced to the application of a specific class of algorithms to data sets and problems, in practice, the successful use of these methods on real-world problems encompasses more than fitting a model to data. In the same way astronomers build data processing pipelines to turn raw data arriving from telescopes into science-ready data products, ML should be understood as an umbrella term not only for a class of algorithms, but also for a set of rules, conventions, and best practices around implementing these models and training them on data. 

The wider scientific community is developing and evolving a set of processes and steps generally implemented as part of a machine learning pipeline \citep[e.g.,][]{kapoor2023reforms}. Best practices within the field of machine learning itself may not always be applicable because the goal in pure ML research is often to find a best-performing model on a given standard training data set. These models can be overly complex and require more ground-truth training data than may be available in realistic situations within astronomy. 

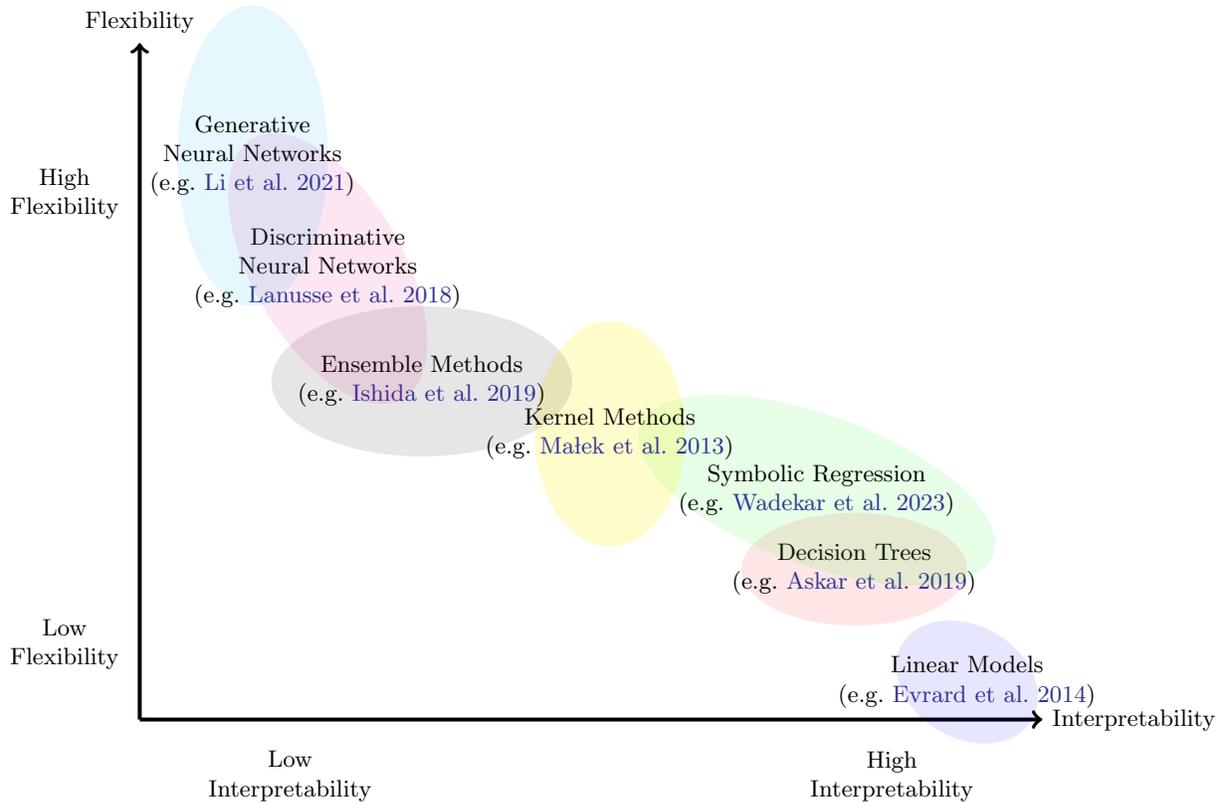
\begin{figure*}
\centering
\begin{tikzpicture}
  \draw[->, line width=1.5pt] (0,0) -- (12,0) node[right] {Interpretability};
  \draw[->, line width=1.5pt] (0,0) -- (0,9) node[above] {Flexibility};
  
  \node[ellipse,fill=blue,inner sep=1pt,label={[align=center]center:Linear Models \\ \citep[e.g.][]{evrard2014model}}, opacity=0.1, minimum width=2cm, minimum height=1.5cm, rotate=-30] (linear) at (11,0.5) {};
  \node[ellipse,fill=red,inner sep=1pt,label={[align=center]center:Decision Trees\\\citep[e.g.][]{askar2019finding}}, opacity=0.1, minimum width=3cm, minimum height=1.5cm, rotate=0] (linear) at (9.5,2) {};
  \node[ellipse,fill=green,inner sep=1pt,label={[align=center]center:Symbolic Regression\\\citep[e.g.][]{wadekar2023augmenting}}, opacity=0.1, minimum width=5cm, minimum height=2cm, rotate=-20] (linear) at (9, 3.05) {};
  \node[ellipse,fill=yellow,inner sep=1pt,label={[align=center]center:Kernel Methods\\\citep[e.g.][]{malek2013vimos}}, opacity=0.2, minimum width=2cm, minimum height=3cm, rotate=-0] (linear) at (6.25,3.8) {};
  \node[ellipse,fill=black,inner sep=1pt,label={[align=center]center:Ensemble Methods\\\citep[e.g.][]{ishida2019optimizing}}, opacity=0.1, minimum width=4cm, minimum height=2cm, rotate=-0] (linear) at (3.75,4.5) {};
  \node[ellipse,fill=magenta,inner sep=1pt,label={[align=center]center:Discriminative\\Neural Networks\\\citep[e.g.][]{lanusse2018cmu}}, opacity=0.1, minimum width=2cm, minimum height=4cm, rotate=30] (linear) at (2.5,6) {};
  \node[ellipse,fill=cyan,inner sep=1pt,label={[align=center]center:Generative\\Neural Networks\\\citep[e.g.][]{li2021ai}}, opacity=0.1, minimum width=2cm, minimum height=4cm, rotate=0] (linear) at (1.5,7.5) {};

  \node[align=center] at (2,-0.75) {Low\\Interpretability};
  \node[align=center] at (10,-0.75) {High\\Interpretability};
  \node[align=center] at (-1,1) {Low\\Flexibility};
  \node[align=center] at (-1,7) {High\\Flexibility};
\end{tikzpicture}
\caption{
Approximate trade-off between flexibility and interpretability of ML models for astronomy. Here, high flexibility generally result in very strong fits to complex training data, but these models may suffer from overfitting, slow computation time, and poor generalization, in addition to non-interpretability. Low flexibility models may be unable to fit extremely complex problems, but will often be the preferred solution in many astronomy problems where simpler baselines have been established. We note that the axes of this plot are qualitative and, in practice, will depend highly on the specific problem. Each referenced model class includes a citation of an example application paper in astronomy.}
\label{fig:tradeoff}
\end{figure*}

In contrast to these highly constrained (and often artificial) problems, in scientific applications, a good rule of thumb is to choose the simplest model possible without sacrificing performance (for a visualization of the trade-offs one might make, see Figure \ref{fig:tradeoff}). Simpler models tend to be more interpretable, and they also facilitate the diagnosis of failure modes.

Beyond this general rule for choosing a model itself, a range of best practices exist for the larger process of ML research: designing and preparing a data set,  choosing evaluation metrics, and exploring and reporting outputs. Common recommendations include splitting the data into training, validation and testing sets, applying cross-validation techniques, scaling input features that span orders of magnitude, using well-calibrated loss functions, and selecting appropriate evaluation metrics.
These practices were generally formulated in order to guard against typical challenges in applying machine learning models, including overfitting and lack of generalization. For example, the performance of the model on data that it has seen during the training phase is not a good estimator of its performance on entirely unseen data, and thus a test set that remains untouched until the model building and validation phase has been completed will give a more informative view of the model's capabilities for generalization. Similarly, evaluation metrics are often noisy estimators of performance, and cross-validation and batch learning may yield more stable estimators of training performance during validation. 

Many tutorials and learning resources utilize standard machine learning data sets such as MNIST. However, these data sets were often specially cleaned and prepared to use when evaluating the performance of a specific machine learning algorithm against other machine learning algorithms. Their properties are generally mismatched with those of data used in real-world applications\footnote{There are current efforts to build a database of benchmarking data sets specifically for astronomy applications of ML algorithms, see \url{https://archive.stsci.edu/hello-universe}}. As a consequence, steps and practices that are crucial in applications of machine learning can be hidden or omitted in learning resources. In order for the astrophysics community to trust a machine learning model, it needs to be convincingly robust and thus should implement best practices aimed at validating model performance. 

For example, \citet{de2022deep} present an application of a Convolutional Neural Network (CNN) to measurements of galaxy cluster masses from \textit{Planck} maps. The authors generate mock images from hydrodynamical simulations processed to include \textit{Planck} observational properties and noise, and use these mock images to train a CNN to learn a relationship between these images and cluster masses. They subsequently apply the trained model to real \textit{Planck} observations and derive cluster masses for 109 galaxy clusters. 
As part of their approach to the problem, they include a thorough analysis of the model and its limitations, including explorations of the flexibility of the model architecture, tests of the stability of the model across multiple training runs, cross-model tests with training data generated with different baryon physics models, as well as theoretical explanations of sources of bias. In doing so, they use best practices as a way of establishing community trust in their methods and results.

Best practices are not laws, and there might be a good reason to use alternative approaches. However, a departure from the standard should be well-motivated and have a clearly articulated reason. For example, some machine learning approaches are embedded in broader contexts, such as simulation-based inference, and thus may benefit from implementing recommendations from adjacent fields like statistics.  Best practices are also neither static nor universal but may depend on the chosen approach and may change with time as our knowledge of machine learning evolves. In some relatively recent areas of development, this evolution is happening rapidly enough that standard recommendations may not yet exist or be in the process of being formulated.
In such cases, researchers should nonetheless aim to establish mechanisms to engender trust in the results of ML models (see also Section \ref{sec:2.3}).

\subsection{Interpret, diagnose, and/or visualize models} \label{sec:2.3}

In many astrophysical applications, using machine learning models as ``black box'' methods is fundamentally at odds with the scientific imperative of understanding nature. It is therefore important to crack open these black boxes and make them at least semi-transparent through the careful use of methods designed to interpret, diagnose, and visualize how a ML model trains and performs on simulated and/or real-world data. This is, in many ways, not very different from non-ML models: both the astronomical community and the field of statistics already apply a wide range of methods to explore model behavior, diagnose model misspecification, and visualize outputs. We strongly encourage researchers to not only apply the existing methods but also to develop new and innovative diagnosis and evaluation techniques to meet the demands and challenges of their research problem. 

There are various ways to monitor the training as well as the in-distribution validation of ML models. One common method is to monitor the training and validation loss as a function of training iteration. These loss curves are helpful for visualizing the optimization procedure and can alert the ML practitioner when training has diverged, has plateaued, or has overfit. 

For regression problems, it is important to track the predicted versus true values. A perfect one-to-one correspondence results in a scatter plot that looks like a line of slope unity. There are many deviations from this ideal performance, such as scatter, bias, and ``regression to the mean'' \citep[for a historical perspective on the concept, see][]{stigler1997regression}.

These sources of error can be diagnosed by carefully selecting validation metrics, such as mean squared error, bias, and/or coefficient of determination ($R^2$). Outlier fraction and outlier-insensitive scatter, such as median absolute deviation, are also important. However, these summary statistics only tell part of the story, as they may strongly depend on the data distribution or saturate too early to differentiate between excellent and suboptimal solutions; it is always recommended to visualize the model output, especially the predicted versus true values of the testing data.

Performance metrics should be selected according to the  problem that is being solved. For example, if a training sample contains 99\% negative classes (non-detections), then any model can trivially attain 99\% accuracy by only predicting the negative class. If the goal of the model is to identify rare detections (the remaining 1\%), then accuracy is a performance metric deeply mismatched with the problem at hand. 
In many applications, it is therefore important to monitor classification problems with the false negative and false positive rates rather than just the error rate (equivalently, check the completeness against purity rather than just accuracy). Usually, classification models depend on a threshold hyperparameter, and by varying this parameter, one can plot the true positive versus false positive rate. This figure is known as the receiving operator characteristic (ROC) curve. 

As an example for a thoughtful approach on diagnosing and interpreting a model, \citet{lapuschkin2019} employ diagnostic methods called layer-wise propagation (LRP) and Spectral Relevance Analysis to explore common image labeling tasks solved by Deep Neural Networks. They identify failure modes in models with very high accuracy, where the network learned, e.g. to classify an image of a horse based on the watermark of the photographer present only on images of horses in the training set. Adding this watermark to images of other subjects such as cars resulted in highly confident changed predictions that the modified image showed a horse when the only change made was this watermark irrelevant to the actual subject shown on the image. Similarly, they found in analyses of the whole dataset that data padding included for technical reasons unwittingly became part of the classification strategy for airplanes, again yielding a model that learned to classify a subject on information irrelevant to the problem. These biases might lead to catastrophic failures when deploying this kind of model in real-world settings. 

ML models have numerous algorithmic failure modes, and blindly trusting their output as a black box will in many cases lead to biased inferences about astrophysical problems. Diagnosing these failure modes while also highlighting where and how the model succeeds, interpreting and visualizing the model's performance on training, and properly using validation and testing data sets are thus of crucial importance in building trust in a model's output. They should be a core part of any data analysis performed with these models.

Modern studies using black-box algorithms often turn to explainability tools as a means of gaining insights into their learned behaviors. Explainable AI \citep{molnar2022} offers a plethora of tools aimed at understanding the decision rules behind black boxes \citep[e.g.][]{ribeiro2016should, lundberg2017unified, selvaraju2022grad}. However, the field is very much in its infancy, and employing such tools for highly complex machine learning models, like deep neural networks, can pose great challenges. To achieve interpretability with such models, a common approach involves establishing an intuitive rationale for why the machine learning model outperforms the baseline, for example, by understanding how flexibility benefits the model or where additional information originates. Subsequently, tests are designed to demonstrate that the models respond to this reasoning when subjected to explainability tools. This process enables a quick assessment of whether a prescribed interpretation is feasible while mitigating confirmation bias and avoiding p-hacking\footnote{The practice of repeated analysis of the same data with multiple hypotheses until a statistically significant result is found, often finding statistically random patterns in data and reporting them as significant.}, which could occur when exhaustively exploring multiple interpretations with minimal effort.

 The explainable AI framework has been criticized for providing at best partial and at worst misleading explanations while legitimizing the reliance on black boxes \citep{rudin2019stop}, based on the argument that a fully faithful explanation would make the original black box model redundant. As an alternative to explainable AI, critics have proposed to rely exclusively on inherently interpretable models \citep[see e.g.][]{chen2019looks}.

\subsection{Explore the limits and scope of models} \label{sec:2.4}

Machine learning models make predictions based on complex and often incompletely understood data. It is crucial to understand the behavior of machine learning models in scientific applications because the consequences of incorrect or biased predictions can have serious implications. While the previous section concerned itself with algorithmic performance (e.g.~overfitting), here we zoom out and consider the limitations of the model more broadly by asking: What data sets can we confidently assume the model will perform well on? And how confident are we that our training data matches our target data? 

The first step in scoping a machine learning application in astronomy is to determine the target problem and the data with which it can be solved.
Models designed to solve practical inference problems with observations \citep[e.g.][]{askar2019finding} should be based on physically observable data. Researchers should be able to explain both how their model inputs can be derived from either idealized or realistic survey observations and also how any limitations in their construction might affect performance. For studies aimed at establishing theoretical limits, such as determination of information content \citep[e.g.][]{de2023robust} or accelerated modeling of simulations \citep[e.g.][]{jamieson2022field}, a more idealized perspective can be adopted by utilizing resources such as the 3D co-moving distribution of galaxies or data from halo finders. Once data requirements are clear, appropriate modeling and testing methodologies can be selected in alignment with the ultimate scientific use case.

Researchers should try to identify where the model's scope stops and to determine which conclusions based on the model are scientifically valid and which are speculation. For example, it is important to estimate the model's predictive power: whether it is purely correlative or whether causal inferences \citep{glymour2019review} can be made. These considerations not only ensure the validity and reliability of a particular research result but also provide clear guidance for reproducing these results and for applications in other research contexts. 

One key limitation of machine learning models in astronomy is that the models can be highly dependent on the quality and quantity of the input data. Applying an ML model to target data outside the scope of the training data is equivalent to asking the model to extrapolate, a task for which most machine learning models are badly suited. For example, \citet{wu2020} showed that different samples' selection criteria resulted in drastically different ML performances; the model was able to extrapolate the relationship between galaxy morphology and galaxy gas content to galaxy voids, but not to galaxy clusters.

If the training data are biased or incomplete, the model may be unable to make accurate predictions on new, unseen data. Complex models can learn the physics behind processes, but as mentioned in the previous rule, they also can learn from unexpected features such as simulation artifacts, and covariant variations. In many applications, machine learning models might be trained on simulated data but applied to real observations. There is a significant danger in these applications because simulating realistic training data is challenging in most practical applications: if we knew the underlying astrophysics perfectly enough to simulate it, we would not need to analyze observations. Similarly, characterizations of telescopes are often incomplete or biased. Combined, these effects might lead to significant covariate shifts between training and target data, which in turn will yield biased results on the target data, even when performance on simulated data is high. For example, the appendix of \citet{ntampaka2022} describes a failure mode they call \textit{overspecialization}: they built an ML model to infer cosmological parameters from multi-wavelength observations of galaxy clusters and identify and found that one model learned details of the specific cosmological simulation it was trained on. While that model performs extremely well on test data generated using this simulation, it would catastrophically fail if applied to  data generated from other simulations or real-world observations. They present strategies to identify and mitigate overspecialization. 

If the model is trained on data from a particular telescope or instrument, it may not generalize to data from other instruments or telescopes \citep{ciprijanovic2023deepastrouda}. Similarly, research implementing ML models trained to emulate physical simulations should clearly state the limits of parameter spaces included in the training data. In this space, \textit{transfer learning} has emerged as a powerful tool to generalize models from specific data sets (e.g.~simulated training data) to a wider range of contexts (e.g.~different target instruments).

The scope of training data is also an issue for use cases where rare, previously unseen events are likely to be part of the target data, and where their identification is part of the classification task. The often serendipitous nature of discovery in astronomy poses a direct challenge to supervised classification approaches, which by definition cannot have seen these unknown phenomena during training. For example, the PLAsTiCC challenge \citep{plasticcdata,plasticc_metrics} explicitly acknowledged and included this by including simulations of currently unobserved, but predicted phenomena in the test data set, but not the training data set. Outlier identification methods have been successfully applied to simulated surveys such as PLAsTiCC \citep[e.g.~][]{ishida2021,villar2021}, and one approach is to build multiple ML models for the same data set, each specialized on a different task (e.g.~outlier detection versus obtaining a pure sample of Type Ia supernovae). Similarly, model diagnostics may include an exploration of model systematics analogous to what is routinely done with numerical models in astrophysics \citep[e.g.~][]{bailin2014} and uncertainty quantification \citep{caldeira2020} derived from statistics.

By understanding the behavior of an ML model, one can identify areas where the model is underperforming. This is a necessary prerequisite to taking steps to improve its performance if part of the scope of a project, for example, by adjusting the model's architecture or by building dedicated models for different prediction tasks. 
Exploring the robustness of the model to covariate shift and to unexplored systematics could involve deliberately introducing systematic differences into the test data set, and subsequently testing the model's performance on these mismatched data sets. For example, testing the model's robustness under changes in the noise properties of the data could help diagnose how the model reacts to mismatches between our parametrization of the telescope response and the real data. Introducing biases into the physical model used for generating training data can explore covariate shifts between training and target data. To understand the model's behavior toward outliers, one may simulate the presence of an unknown phenomenon in the target data by holding out an entire rare, but known, class from the training data. 

Ideally, model predictions should be fair, transparent, and unbiased. In many situations where training data may be imperfect or incompletely understood, it may be impossible to build unbiased models. In these situations, transparency is thus of crucial importance in order to be able to apply results in ways that limit biases in astrophysical interpretation. The failure modes of the model should be faithfully described.
If a model is susceptible to sources of non-physical error, such as simulation artifacts or covariate shift, it must be tested against these to prove it is robust. This is particularly relevant for methodologies with surprising results.
If these tests are beyond the scope, it should be very clear to readers the cases in which the model might fail, and how biases in the predictions might propagate to higher-level uses of the model's outputs.

\subsection{Share and discuss the lessons learned} \label{sec:2.5}

Machine learning in astronomy is in many ways exploratory: as mentioned above, ML methods are  applied to complex data sets to answer equally complex questions and often require significant adaptation in order to yield answers to astronomy's most interesting questions. This process is nonlinear, and includes successes as well as informative failures. At the same time, publications are biased towards discussing successes \citep[e.g.~][]{liddle2004,vaughan2008,foley2012,degrijs2014,jaffarian2020,tiede2020}, often the end points of years-long explorations of methods, architectures, activation functions, and performance metrics. Some models may never get published for not reaching the thresholds of qualitative or quantitative improvement over previous approaches. As with other cases of publication bias, this does the community a disservice in the long run, because it obscures important information about where and how models can (or cannot) be applied to astronomical data, and it ultimately may lead to research groups repeating other groups' futile attempts. The growing corpus of research results applying ML in astronomy necessitates writing manuscripts that are pedagogical as well as scientific to inform the next generation of researchers, and this teaching is best done with examples of both successes and failures.

In many ways, this rule is a direct result of the careful design and exploration we advocate for above: failures might not be evident without significant model exploration, and similarly, catastrophic successes can sometimes actually be failures in disguise \citep[e.g.,][]{lapuschkin2019}. Biasing publication exclusively towards successful approaches could discourage model exploration to avoid the risk of failure, while at the same time depriving the literature of these examples, which still have the capacity to inform research directions on related problems.

As an example, researchers might include alternative, failed approaches to the problem in the appendix of a publication, which would provide the relevant information without cluttering the main text \citep[for an example, see e.g.~][]{ntampaka2022}. Similarly, there is a growing recognition of the importance of null results and failed approaches within the ML community as well. For example, ``I Can't Believe It's Not Better!''\footnote{\url{https://i-cant-believe-its-not-better.github.io/}} is a session at the NeurIPS ML Conference that is aimed at showcasing meaningful research beyond impressive performance metrics, rewarding approaches beyond those that beat previous works on benchmarks. As we argue in Rule 2 above, common performance metrics used in the ML community, executed on curated example data sets, are often mismatched to real-world applications in astronomy. As a community, there is value in aligning ourselves with and extending initiatives that look beyond simple metrics and report a wider range of results. 

\subsection{Make software and data publicly available} \label{sec:2.6}

Public code and public data enable others to reproduce research results and verify correctness, and provide a foundation for new projects. While many machine learning projects may be designed with a narrow focus on solving a particular problem, there is a growing recognition that model architectures, loss functions, and indeed sometimes whole pipelines transfer well to problems that concern a very different area of astrophysics but where the underlying structure of the problem is analogous. Making data, code, and models public not only ensures reproducibility but also encourages reuse. 

As we explain in the previous rules, new tools and methods should be benchmarked against existing models (see e.g.~\citealt{lueckmann2021} for an instructive example on benchmarking, including an interactive website\footnote{\url{https://sbi-benchmark.github.io}} and software to benchmark and compare new methods as they are developed), and these should be published even when a model might not reach current benchmarks. 
These comparisons are best done using or closely following the original code. Replicating someone else's code is at best time-consuming and error-prone, and at worst impossible.
Authors will benefit from the code of their colleagues, and in turn, colleagues will benefit from your code when they need to compare their own new tools. In the context of closed collaborations, data, and model sharing rights may be complicated, and sharing the full data set or model may be prohibited. In these cases, \textit{toy} data and code are valuable substitutes. 

There are many good options for sharing code and data. \href{https://github.com}{GitHub} is currently the de facto standard for archiving public code, but alternatives exist (e.g. \href{https://about.gitlab.com/}{GitLab}, \href{https://bitbucket.org/product/}{Bitbucket}).
It is important to realize that a code archive is likely to last longer than software dependencies and that breaking changes in future versions of those dependencies will make running a specific code base increasingly difficult.
The rapid pace of ML framework development in particular makes reusing old models complicated; models may not load correctly into newer environments.
Active maintenance of code and models is recommended when this code generalizes to a wide range of problems, or in cases where the authors anticipate that the software will become a widely used tool. In the absence of that requirement, we recommend specifying the exact requirements for running your code (e.g. by exporting a virtual environment, or with \texttt{pip freeze}) or by creating a Docker image.
\href{https://zenodo.org}{Zenodo} is a free service funded by the \mbox{OpenAIRE} project for preserving and sharing small volumes of code and data. It integrates with GitHub to version and to assign unique Digital Object Identifiers (DOIs) to both code and data, efficiently enabling reproducibility in many contexts.

Public code is not the same as user-friendly code. Ideally, machine learning code should be designed with an accessible API and accompanied by documentation. But making research code easy to use takes time, and writing popular software is sometimes poorly recognized in comparison to traditional measures of productivity\footnote{As a workaround, the Journal of Open Source Software (JOSS) publishes short papers on research software.}. We encourage researchers to choose the degree of useability that best suits their research and goals. As a starting point, they might consider their audience; if code is only intended for a few experienced machine-learning-focused astronomers, clean code and docstrings may be enough, but if the goal is to produce a tool that is widely adopted, more refactoring and documentation (and hence more development and maintenance time) may be required.

\section{Rules are Meant to be Broken} 
\label{sec:breaking}

While we have named them rules, the guidelines set out in the previous sections should not be seen as absolutely prescriptive in every case. We have attempted to lay out current best practices in a rapidly changing field, and we know from our own research experience that not every rule applies to every situation and every model. Rather than compiling a comprehensive guide, we have collected key features that we believe will make machine learning research in astronomy more robust, trustworthy, and accurate. Rather than considering the rules above as the one prescriptive way to a well-designed ML project, consider them an encouragement to engage critically and deeply with the methods being used, explore the project's limitations and evaluate its performance with a view toward interpretability and usefulness in astrophysical contexts. This is equally true when embarking on a machine learning research project and when reviewing both proposals for and the outputs of such a project, including publications and software.

We expect that there are many among the growing community of experts who will disagree with some or all of the rules above \textemdash{} even among authors of this manuscript, there is not complete consensus on every rule, and it would be easy to find examples among our own publications and projects that break several of them! We consider this need for nuance as a feature, not a bug. Critically engaging with the rules is a fruitful way to develop a set of best principles. There are exceptions to all rules, but making well-motivated exceptions requires knowledge of what the rules are, and why they are important. Deviations from our recommendations are expected and indeed encouraged. But as we have tried above to be specific and well-reasoned in our motivations above, we expect projects implementing different approaches to do the same. Our advice to researchers new to machine learning is to start with the recommendations above and work from there, adapting them to their own specific context where required.

Because this manuscript focuses on the application of ML to astronomical data, we have largely left out ethical and privacy considerations from the rules above. We note, however, that we develop and apply ML in the context of the larger society we live in, and that ethical and privacy considerations are important when training students in ML methods, when applying models to data generated by or about humans, or when decisions made based on ML models affect humans (for an overview of salient debates and guidelines, see also \citealt{jobin2019, sep-ethics-ai}). 

\section{Conclusion}

Machine learning is a powerful tool for astrophysical discovery, and it is rapidly becoming an indispensable tool in modern astronomy. There are many applications where machine learning can lead to important scientific advances. At the same time, like any methodology, ML methods come with requirements, limitations, and trade-offs compared to other approaches. We encourage researchers to adopt a critical and thoughtful approach to ML research projects and their review. The rules here are meant to be a starting point, and as the field matures, we anticipate they might undergo further evolution and refinement. 

\acknowledgements{}
We thank Annalisa Pillepich and all the organizers and attendees of the Ringberg meeting on ``Machine
Learning Tools for Research in Astronomy 2019''
(\url{https://www2.mpia-hd.mpg.de/ml2019/}) for the productive
discussions that led to the plan for this manuscript.
This project was developed in part at the 2022 Astro Hack Week, hosted by the Max Planck Institute for Astronomy  and Haus der Astronomie in Heidelberg, Germany.
This work was partially supported by the Max Planck Institute for Astronomy, the European Space Agency, the Gordon and Betty Moore Foundation, and the Alfred P. Sloan Foundation.  M.N. is supported in part by NASA under award No. 80NSSC22K0821.
D.H. is supported by the Women In Science Excel (WISE) programme of the Netherlands Organisation for Scientific Research (NWO). M.H. is supported by the Simons Collaboration on Learning the Universe.  M.P. acknowledges financial support from the European Union’s Horizon 2020 research and innovation program under the Marie Skłodowska-Curie grant agreement No. 896248.  N.R.’s work at Argonne National Laboratory was supported under the U.S. Department of Energy contract DE-AC02-06CH11357. M.W. acknowledges funding from the Science and Technology Facilities Council (STFC) Grant Code ST/R505006/1.

\bibliography{references}

\clearpage
\appendix

\section{A Quick-Start Guide for Assessing ML Astronomy Research}
\label{sec:appendix}

Interdisciplinary results can be difficult to assess because they require a deep understanding not only of the scientific domain, but also of the methodology.  This quick-start guide is intended as a starting point for readers and referees to assess new research for which they have a domain understanding but may lack methodological context. For referees, we caution that this guide is not absolutely prescriptive, nor exhaustive. Referees should consult journal expectations \citep[e.g.][]{AAS-ref, MNRAS-ref, Nature-ref} and more general refereeing references \citep{survive-peer,  nicholas2011quick, Raff-ref, 2022BAAS...54..051N} for guidelines on best practices for providing evaluations of manuscripts. The following considerations do not replace refereeing best practices; instead, they are additions to best practices that are specific to evaluating ML astronomy research.


\begin{enumerate}
    \item Compare against a domain reference and put results in the broader context.
    \begin{enumerate}
        \item Are the results put in the appropriate context?  For example, if the method replaces an existing ``traditional'' technique, are results (accuracy, compute cost, robustness, etc.) from the traditional technique used as a comparison?
        \item In some cases, the new ML method enables an analysis that was not possible before; no traditional benchmark exists.  In this case, it may be more difficult to put the results in context.
        \item If the outputs of the model are likely to be used downstream (e.g.~for population-level analyses), do the authors consider how biases in their model might propagate into these analyses?
    \end{enumerate}
    \item Adopt best practices from the ML community.
    \begin{enumerate}
        \item Have the authors included citations in their literature review to summarize particular best practices applied in their work?
        \item Have the authors used current best practices?  If they have not, have they justified this choice in the context of the scope of their work?
    \end{enumerate}
    \item Interpret, diagnose, and/or visualize models.
    \begin{enumerate}
        \item Does the research include interpretation, diagnosis, or visualization of the model?
        \item Are algorithmic failure modes being explored?
        \item Does the discussion attempt to explain the interpretation in light of the physical system that is being studied?
    \end{enumerate}
    \item Explore the limits and scope of models.
    \begin{enumerate}
        \item Have the authors explored and identified the scope of the training data and the model?
        \item Does the research explore how the model responds to perturbations?  These might include a change in simulation modeling technique, the introduction of noise, or modeling new instrumental effects.
        \item Does the research put the model in the appropriate context of these results, e.g., by discussing cases where the model can be safely used and also cases where the model may break down?
    \end{enumerate}
    \item Share and discuss the lessons learned.
    \begin{enumerate}
        \item Does the manuscript include information about the research process by, for example, giving results from other models or techniques that were explored and abandoned?  Does the manuscript explain why those techniques were abandoned?
        \item If the authors' research resulted in informative failures, does the manuscript discuss and attempt to explain these failures?
    \end{enumerate}
    \item Make software and data publicly available.
    \begin{enumerate}
        \item Is the data set used in the paper available?  Note that in some cases, data may be a subset of a larger corpus hosted elsewhere, it may be proprietary, or it otherwise may be difficult to share, and in this case, toy data may be a reasonable alternative.
        \item Is the software available?  While some applications benefit from releasing a comprehensive, end-to-end, maintained software pipeline, toy code is sufficient for others. The use case should be carefully considered.
    \end{enumerate}
\end{enumerate}

\end{document}